\documentstyle{article}
\begin{document}
\title{\hspace{5mm}Low Energy Behaviour of XXZ Antiferromagnetic Spin
Chain}
\author{B. Roostaei\footnote{Email:Bahmanr@mehr.sharif.ac.ir}}
\date{\small \it Department of Physics,Sharif University of Technology,P.O. Box 19395-5534,Tehran-Iran \\
and Institute for Studies in Theoretical Physics and
Mathematics(IPM),P.O. Box 11365-9161,Tehran-Iran} %
\maketitle %
\centerline{{\bf Abstract}}%
{\small The zero temperature phase diagram of XXZ spin chain in
external magnetic field is investigated at low energies using
path integral approach.It has been shown by spin wave analysis
and then by nonlinear sigma model transformation that below some
critical field the system undergoes quantum Kosterlitz-Thouless
phase transition.Above that critical field for low anisotropies
the system is saturated while at strong anisotropy the system has
Ising antiferromagnetic order.}
\section{Introduction}
 Considerable effort has been made to understand the long
range (low energy) behaviour of quantum spin systems specially in
low dimensions.The importance of these studies is in their
applicability in various physical phenomena in for example quantum
Hall ferromagnets[1], high $T_c$ superconductors[2] and many
other magnetic systems.On the other hand we know very few models
to be exactly solvable, nevertheless development of field
theoretical approaches has greatly changed the language in this
part of strongly correlated systems and has provided us with
effective methods to be applied to our problems of spin
systems.One of the great observations by means of field
theoretical approach has been the problem of ground state of 1D
Heisenberg antiferromagnet: after separating out the rapidly
varying fluctuations we obtain a nonlinear sigma
model(NL$\sigma$M) with a topological term which is responsible
for destroying the long range order of ground state for integer
spin chains according to Haldane's conjecture[3] while we know
that this problem is exactly solvable by Bethe's ansatz only for
spin half problem[4].The Haldane's conjecture has been now
confirmed by strong experimental and numerical evidence[4,5,6].
The method of separating out short wavelength fluctuations which
resulted in NL$\sigma$M+topological term for antiferromegnetic
spin chains is now widely used for various spin systems including
Ferrimagnets[7] and spin ladders[8].In this paper I am going to
apply this method of obtaining low energy behaviour to anisotropic
spin chains.
\section{Spin Wave Analysis}
For the anisotropic Heisenberg model in magnetic field:
\begin{equation}
\hat{\mathcal {H}}=J\sum_{<ij>} [{1\over 2}
(S_i^{+}S_j^{-}+S_i^{-}S_j^{+})+\lambda
S_i^{z}S_j^{z}]+\sum_{j=1}^{N}{\bf B}\cdot{\bf S}
\end{equation}
where $0\leq\lambda\leq 1$ there are numerical evidence[9] that
under some critical value for the magnetic field $B_c$ the $S_z$
fluctuations become negligible and the system behaves as XY model
in long range.On the other hand we know that the isotropic
Heisenberg antiferromagnet becomes massive by turning on magnetic
field[8].These two considerations suggest that $B_c\propto
(1-\lambda)^{-\nu}$ for XXZ model.This may be examined by some
crude spin wave analysis: I Apply the Holstein-Primakov
transformation which is some kind of perturbation around
saturation point: $S_i^{z}=s-a_i^{\dag}a_i$ and
$S_i^{-}=(2s)^{1\over 2}a_i^{\dag}(1-{a_i^{\dag}a_i\over
2s})^{1\over 2}$. Here $a^{\dag}$ creates a boson to lower the
$S^{z}$ by one.We are in a situation where the number of bosons
are low enough to approximate the square root and obtain:
$S_i^{-}\approx (2s)^{1\over 2}a_i^{\dag}$.Applying this
transformation to the Hamiltonian (1) for ${\bf B}=B\hat z$ and
Fourier expanding we find:
\begin{equation}
\hat{\mathcal{H}}\approx
\sum_{k}[2(\gamma_k-z\lambda)Js-B]a_k^{\dag}a_k
\end{equation}
in which $\gamma_k\equiv\sum_{\bf a}e^{ik\cdot {\bf
a}}=\gamma_{-k}$. ${\bf a}$ represents the translation vectors to
the neighboring sites and $z$ is the number of nearest neighbors.
\newline
In one dimension the spin wave spectrum becomes gappful:
$E_k=4Js(\cos ak-\lambda)-B$ with the gap $\Delta
=4Js(1-\lambda)-B$.This estimation confirms my suggestion and sets
$\nu =1$.Under $B_c$ the saturation point becomes unstable.This
elementary analysis will be invalid as $B$ approaches zero and is
valid just near $B_c$.This result shows that near $B_c$ the
correlation length of $S_z$ fluctuations $\xi \propto \mid
B-B_c\mid^{-1}$. In section 3 after separating out short range
$Ne\grave{e}l$ field we can see that the correlation length of
$S_z$ fluctuations is actually $\xi\propto (1-\lambda)^{-{1\over
2}}$ for $\lambda\leq 1$.
\newline
For the case of $\lambda\geq 1$ the analysis is a bit
complicated.In this regime to find the fluctuations around Ising
antiferromagnetic order(IAF) we can again apply the
Holstein-Primakov transformation for two sublattices:
$S_i^z=s-a_i^{\dag}a_i$ and
$S_i^{-}=\sqrt{2s}a_i^{\dag}(1-{a_i^{\dag}a_i\over 2s})^{1\over
2}$ for $i$ in sublattice A and for sublattice B:
$S_i^z=-s+b_i^{\dag}b_i$ and
$S_i^{-}=\sqrt{2s}(1-{b_i^{\dag}b_i\over 2s})^{1\over 2}b_i$.
Here we have two kind of Bosons one $a^{\dag}(b^{\dag})$
lowers(adds) the $S_z$ by one in sublattice A(B).By applying the
first order approximation: $S_i^{-}\approx \sqrt{2s}a_i^{\dag}$
for A and $S_i^{-}\approx \sqrt{2s}b_i$ for B, and inserting in
the Hamiltonian (1) we find:
\begin{eqnarray}
\hat{\mathcal{H}}\approx Js\sum_{i\in A}\sum_{i\in
B}[a_ib_j+a_i^{\dag}b_j^{\dag}+\lambda
(a_i^{\dag}a_i+b_j^{\dag}b_j)]+B\sum_{i\in
B}b_i^{\dag}b_i-B\sum_{i\in A}a_i^{\dag}a_i \\ \nonumber %
=Js\sum_k \gamma_k(a_kb_{-k}+a_k^{\dag}b_{-k}^{\dag})+\alpha\sum_k
b_k^{\dag}b_k+\beta\sum_k a_k^{\dag}a_k \; . \hspace{1.8cm}
\end{eqnarray}
Where $\alpha\equiv Js\lambda z+B$ and $\beta\equiv Js\lambda
z-B$ and we have Fourier expanded the operators in the last
equality . Now to diagonalize (3)I apply the (canonical)
Bogoliubov transformation :
\begin{eqnarray}
c_k=u_ka_k-v_kb_{-k}^{\dag} \\ \nonumber %
d_k=u_kb_k-v_ka_{-k}^{\dag}
\end{eqnarray}
For the transformation to be canonical we must have:$u^2-v^2=1$
and for (3) to be diagonal we must have:
$Js\gamma(u^2+v^2)+(\alpha+\beta)uv=0$. No problem arises if we
assume $u$ and $v$ are real. Finally we obtain:
\begin{eqnarray}
\hat{\mathcal{H}}\approx\sum_k {\mathcal{E}}^{(1)}_k
c_k^{\dag}c_k+\sum_k {\mathcal{E}}^{(2)}_k d_k^{\dag}d_k  \\
\nonumber {\mathcal{E}}^{(1,2)}_k\equiv
2Js(\lambda^2-\cos^2ak)^{1\over 2}\pm B .
\end{eqnarray}
One of The branches is always gapfull which has to be the
spectrum of fluctuations trying to align spins against the
magnetic field direction and the other one is also gapfull which
tries to saturate the system, but in this branch the gap vanishes
at: $B_c=2Js(\lambda^2-1)^{1\over 2}$. This result shows that for
$\lambda\geq 1$ the correlation length $\xi\propto
(\lambda-1)^{-{1\over 2}}$ that is $\nu ={1\over 2}$ above the
critical point $\lambda =1$. This result is more reliable for the
following reason:
\newline
we can compute the site magnetization reduction due to quantum
fluctuations in our spin wave approximation:
\begin{eqnarray}
\mid\Delta <S_z>\mid = \int_{-{\pi\over a}}^{\pi\over a} v_k^2 dk
=
 \int_{-{\pi\over a}}^{\pi\over a} (\frac{\lambda z/2}{\sqrt
 \lambda^2z^2-\gamma_k^2}-{1\over 2})dk \\ \nonumber
 =-{2\pi\over a}+{4{\sqrt 2}\lambda\over a}\frac{F({-1\over
 \lambda^2-1})}{\sqrt{\lambda^2-1}} \; .
 \end{eqnarray}
Where $F$ is the complete elliptic integral of first kind.We can
see in Fig.I that this reduction is finite at $\lambda > 1$ and
vanishes very fast as $\lambda$ grows. %
\par
Now we can sketch the line of transition from saturation to IAF
and other phases as follows(fig.II):
\begin{eqnarray}
B_c(\lambda)=2Js\sqrt{\lambda^2-1} \hspace{1cm} \lambda\geq 1
\\ \nonumber
B_c(\lambda)=4Js(1-\lambda) \hspace{1.2cm} \lambda\leq 1   .
\end{eqnarray}
\section{Field Theoretic Discussion}
In this section I am going to analyze the anisotropic spin chain
by means of nonlinear sigma model transformation. The Minkowskian
action for anisotropic Heisenberg chain
${\hat{\mathcal{H}}}=\sum_{j=1}^{N}\alpha_{\mu\nu}S_j^{\mu}S_{j+1}^{\nu}-\sum_{j=1}^{N}{\bf
B}\cdot {\bf S}$ with the symmetric rank two tensor $\alpha$ is :
\begin{equation}
S_M=s\sum_{j=1}^{N} S_{WZ}[{\bf n}_j]-\int_{0}^{\beta\hbar} dx_0
\sum_{j=1}^{N}
\alpha_{\mu\nu}n_j^{\mu}n_{j+1}^{\nu}-\int_{0}^{\beta\hbar}dx_0
\sum_{j=1}^{N} {\bf B}\cdot {\bf n} \; .
\end{equation}
Where $x_0$ is the real time and $\beta=1/k_BT$ that in this
paper we discuss the zero temperature situation
$\beta\hbar\rightarrow\infty$.The first term known as Wess-Zumino
term is a Berry phase for each spin and represents the area of a
cap of the unit sphere formed by the closed path passed by the
unit vector ${\bf n}(x_0)$ from $x_0=0$ to $x_0=\beta\hbar$[10]
.Under the critical magnetic field $B_c$ we can separate out
rapidly varying fluctuations as follows:
\newline
First of all we stagger the field: ${\bf n}_j\rightarrow
(-1)^j{\bf n}_j$ then we write ${\bf n}_j$ as a short range
$Ne\grave{e}l$ field ${\bf l}_j$ plus fluctuations around it
${\bf m}_j$ : ${\bf n}_j={\bf m}_j+a_0(-1)^j{\bf l}_j$ where
${\bf m}^2=1$, ${\bf m}\cdot {\bf l}=0$ and $a_0$ is the lattice
constant. This method works only for two sublattice systems close
to $Ne\grave{e}l$ state which is not the eigenstate of
$\hat{\mathcal{H}}$ of course!
\newline
Expanding the action up to second order in ${\bf l}$ and
integrating it out we obtain $ Z=\int {\mathcal{D}}{\bf m}\delta
({\bf m}^2-1) \; e^{iS_{eff}}$ where :
\begin{eqnarray}
S_{eff}=\int d^2x [{1\over
a_0}\alpha_{\mu\nu}m^{\mu}m^{\nu}+{a_0\over
 2}\alpha_{\mu\nu}m^{\mu}\partial_1^2m^{\nu}]+
 {s\over 2}\int d^2x \; {\bf m}\cdot (\partial_0{\bf m}\times
\partial_1{\bf m})+ \\ \nonumber
+{s^2\over 4a_0}\int\alpha_{\mu\nu}^{-1}[({\bf m}\times\partial_0
{\bf m})_{\mu}({\bf m}\times\partial_0 {\bf m})_{\nu}]d^2x
-{1\over Js\lambda a_0}{\bf B}\cdot ({\bf m\times\partial_0 m}) .
\hspace{0.8cm}
\end{eqnarray}
For the XXZ model $\alpha_{\mu\nu}=Js^2 \; diag(1,1,\lambda)$ by
which we find:
\begin{eqnarray}
{\mathcal{L}}_{eff}=\frac{s}{4a_0 J}(\partial_0
m_z)^2-{a_0J\lambda\over 2}(\partial_1 m_z)^2-{J\over
a_0}(1-\lambda)m_z^2-{a_0J\over 2}[(\partial_1 m_x)^2+ \\
\nonumber +(\partial_1 m_y)^2]+\frac{1}{4a_0J}\{(\partial_0
m_y)^2[m_z^2+\lambda^{-1}m_x^2]+(\partial_0
m_x)^2[m_z^2+\lambda^{-1}m_y^2]+ \\ \nonumber
-2(m_ym_z\partial_0m_z\partial_0m_y+m_xm_z\partial_0m_x\partial_0m_z+\lambda^{-1}m_xm_y\partial_0m_y\partial_0m_x)\}+
\\ \nonumber
-{B\over Js\lambda a_0}({\bf m\times\partial_0 m})\cdot\hat{z}
+{\theta\over 8\pi}\epsilon_{\mu\nu}{\bf
m}\cdot(\partial_{\mu}{\bf m}\times\partial_{\nu}{\bf m})\; .
\hspace{3cm}
\end{eqnarray}
where for the last term $\mu ,\nu =0,1$.The last term is the
Pontryagin index (or winding number) of field ${\bf m}$ as a map
from $S^2$ (real space-time) to $S^2$ (configuration space) and
is a topological term and $\theta=2\pi s$[10]. This is the well
known $\theta$-term that for isotropic Heisenberg chain resulted
in Haldane's conjecture, however in XXZ chain as we can see in
(10) due to massive fluctuations of $m_z$ $(\Delta\propto
1-\lambda)$ the $z$ component will be negligible in ground state,
in other words this component is short range and so at long
wavelengths(low energies)is irrelevant.To find the low energy
behavior we can use the mean-field approximation $m_z\approx 0$.
In this approximation the winding number will be ineffective.
\section{Topological Phase Transition at T=0}
In our mean-field approximation removing the $z$ component in the
Lagrangian density results in:
\begin{eqnarray}
{\mathcal{L}}_{eff}=-{a_0J\over 2}[(\partial_1 m_x)^2+(\partial_1
m_y)^2]+\frac{1}{4a_0 J\lambda}[(\partial_0 m_y)^2m_x^2+ \hspace{2cm} \\
 \nonumber
+(\partial_0 m_x)^2m_y^2 -2m_xm_y\partial_0 m_x\partial_0 m_y]-\frac{B}{J\lambda sa_0}(m_x\partial_0m_y-m_y\partial_0m_x) \\
\nonumber
\rightarrow\frac{1}{2g}[(\partial_0\varphi)^2-(\partial_1\varphi)^2]-{2B\over
g s}(\partial_0 \varphi) \; . \hspace{4.3cm}
\end{eqnarray}
Where in the last line we have parametrized ${\bf m}$ as
$(\cos\varphi,\sin\varphi,0)$ and scaled the space-time so that
the spin wave velocity is one and $g=(2\lambda)^{1\over 2}$ is
the coupling constant.This is the Lagrangian for massless
fluctuations of XXZ model in 1+1 dimensions.(The term coupled to
magnetic field is a total derivative, i.e. the magnetic field has
no effect on in plane excitations in mean-field approximation.)
The Lagrangian in (11) has topological defects for the field
$\varphi(x_0,x)$ is an angle.These defects drastically alter the
behavior of system when they are present.In this
scenario(Kosterlitz-Thouless), at weak couplings $(g<g_c)$ the
ground state has bound vortices(defects in space-time) and is
partially ordered with algebraically falling correlations.At
strong couplings the vortices become unbound and the partial order
completely vanishes and the correlations fall off exponentially
with correlation length $\xi\propto 1/\ln(2g))$.This drastic
change is a second order quantum phase transition at zero
temperature without symmetry breaking.
\section{Summary}
The low energy phase diagram of antiferromagnetic XXZ chain in
magnetic field is shown in fig.II . At $0\leq\lambda<1$ and under
$B_c\approx 4Js(1-\lambda)$ we have two kind of ground states: at
$\lambda<\lambda_{KT}$ the ground state is partially ordered while
at $\lambda>\lambda_{KT}$ the ground state has free vortices and
is disordered.Above the critical magnetic field the $z$ component
is no longer ignored and we have saturated ferromagnetic(SF)
order.At $\lambda>1$ and $B<B_c$ obviously the system has
antiferromagnetic Ising (IAF) behaviour at low energies while in
this regime and above $B_c$ we have again saturated phase. For
$B=0$ it is inferred that we have a quantum phase transition at
$\lambda=1$ from disordered phase to ordered IAF phase which has
already been shown in continum limit for spin half case by
Bosonization method[10].
\section{Acknowledgement}
The author is grateful to IPM for funding this research as a
student research work .
\section{References}
[1]S.M. Girvin, Cond-Mat/9711233 and references therein. \newline
[2]P.W. Anderson, Science{\bf 235},1196(1987), S.Chakravarty,B.I.
Halperin,D.R. Nelson, Phys. Rev. B,{\bf 39},2344(1989) and
references therein. \newline [3]F.D.M. Haldane, Phys. Rev.
Lett.,{\bf 50},1153(1983). \newline [4]E. Fradkin,{\it Field
theories of Condensed Matter Systems}(Addison-Wesely,1991),Chp.4
\newline [5]I. Affleck, J. Phys. :Cond. Matt. {\bf
1},3047(1989). \newline [6] A. Auerbach, {\it Interacting
Electrons and Quantum Magnetism}(Springer-Verlag,New York,1995) .
\newline
[7]R. Abolfath, H. Hamidian,A. Langari, Cond-Matt/9901063 .
\newline
[8] B. Normand,J. Kyriakidis,D. Loss, Cond-Matt/9902104 .
\newline
[9]A. Langari, Phys. Rev.B,{\bf 58},14467(1998) .
\newline
[10]Ref.4 charter 5.
\newline
[11]A.Lopez,A.G.Rojo,E.Fradkin, Phys. Rev. B,{\bf 49},15139(1994).
\vspace{2cm}
\begin{center}
Figure caption
\end{center}
Fig. I)Site magnetization reduction due to quantum fluctuations
at ground state in the regime of higher anisotropy $\lambda > 1$ .
\newline
\newline
Fig. II)The low energy phase diagram of the XXZ antiferromagnetic
chain.The straight line shows the threshold of saturation.The
curve $\sqrt{B^2+1}$ is the threshold of Ising
Antiferromagnetism(IAF).
\end{document}